\documentclass[11pt]{article}\pdfoutput=1
\usepackage{cite}
\usepackage{amsmath,amsfonts,amssymb}
\usepackage[small,bf,hang]{caption}
\usepackage{slashed}
\usepackage{amsmath}
\usepackage{latexsym,epsfig}
\usepackage{arydshln}

\def\hybrid{
        \topmargin -20pt
        \oddsidemargin 0pt
        \headheight 0pt \headsep 0pt
        \textwidth 6.25in 
        \textheight 9.5in 
        \marginparwidth .875in
        \parskip 5pt plus 1pt \jot = 1.5ex}

\hybrid

 \csname
@addtoreset\endcsname{equation}{section}

\def\moth{\mathsurround=0pt}
\newdimen\zo \zo=0pt

\def\tick{\leaders\hrule height 0.5ex depth 0pt \hskip 0.5pt}
\def\upboxfill{$\moth \setbox\zo\hbox{\tick}%
  \hskip 3pt\hbox to 0pt{$\tick$\hss}\hrulefill \hbox to 7.5pt{$\tick$\hss}$}

\def\dtick{\leaders\hrule height .34pt depth 0.5ex \hskip 0.5pt}
\def\downboxfill{$\moth \setbox\zo\hbox{\dtick}%
  \hskip 2pt\hbox to 0pt{$\dtick$\hss}\hrulefill \hbox to 2pt{$\dtick$\hss}$}

\def\bec{\begin{center}}
\def\ec{\end{center}}

\def\be{\begin{equation}}
\def\ee{\end{equation}}
\def\bea{\begin{eqnarray}}
\def\eea{\end{eqnarray}}
\def\ba{\begin{array}}
\def\ea{\end{array}}

\begin{document}

\begin{titlepage}
\rightline{}
\rightline{September 2017}
\rightline{MPP-2017-209}
\rightline{LMU-ASC 57/17}
\begin{center}
\vskip 1cm
{\Large \bf{Constructions of L$_{\infty}$ algebras \\[1ex]
and their field theory realizations}
}\\
\vskip 0.5cm


  \vskip 1.0cm
 {\large \bf {Olaf Hohm$^{1}$, Vladislav Kupriyanov$^{2,3}$, Dieter L\"ust$^{2,4}$, Matthias Traube$^{2}$}}
\vskip 1cm

{\em  \hskip -.1truecm $^1$Simons Center for Geometry and Physics, \\
Stony Brook University, \\
Stony Brook, NY 11794-3636, USA \vskip 5pt }

{\em $^{2}$ Max-Planck-Institut f\"ur Physik,
  Werner-Heisenberg-Institut\\ F\"ohringer Ring 6, 80805 M\"unchen, Germany
\vskip 5pt}

{\em $^{3}$  Universidade Federal do ABC}\\
{\it Santo Andr\'e, SP, 
Brazil
\vskip 5pt}

{\em $^{4}$ Arnold Sommerfeld Center for Theoretical
  Physics, Department f\"ur Physik\\ Ludwig-Maximilians-Universit\"at
  M\"unchen\\ Theresienstra{\ss}e 37, 80333 M\"unchen, Germany\vskip 5pt}

ohohm@scgp.stonybrook.edu, vladislav.kupriyanov@gmail.com, dieter.luest@lmu.de, mtraube@mpp.mpg.de\\

\vskip 1cm
{\bf Abstract}
\end{center}


\noindent
\begin{narrower}

We construct L$_{\infty}$ algebras for general `initial data' given by 
a vector space equipped with an antisymmetric bracket 
not necessarily satisfying the Jacobi identity.   
We prove that any such bracket can be 
extended to a 2-term L$_{\infty}$ algebra 
on a graded vector space of twice the dimension, with the 3-bracket being related to the Jacobiator. 
While these L$_{\infty}$ algebras always exist, they generally do not realize a non-trivial 
symmetry in a field theory. 
In order to define L$_{\infty}$ algebras with genuine field theory realizations, 
we prove the significantly more general theorem that if the Jacobiator takes 
values in the image of any linear map that defines an ideal  
there is a 3-term 
L$_{\infty}$ algebra with a generally non-trivial 4-bracket. 
We discuss special cases such as the commutator algebra of octonions, 
its contraction to the `R-flux algebra', and the Courant algebroid.

\end{narrower}

\end{titlepage}

\tableofcontents

\section{Introduction}

Lie groups are ubiquitous in mathematics and theoretical physics as 
the structures formalizing  the notion of continuous symmetries. 
Their infinitesimal objects are Lie algebras: 
vector spaces equipped with an antisymmetric bracket  
satisfying the Jacobi identity. In various contexts it is advantageous (if not strictly required)  
to generalize the notion of a Lie algebra so that the brackets do not satisfy 
the Jacobi identity. Rather, in addition to the `2-bracket', 
general `$n$-brackets' $\ell_n$ are introduced on a graded vector space 
for  $n=1,2,3,\ldots$, satisfying generalized Jacobi identities 
involving all brackets. 
Such structures, referred to as L$_{\infty}$ or strongly homotopy Lie algebras, first appeared in the physics literature 
in closed string field theory \cite{Zwiebach:1992ie} and in the mathematics literature in topology \cite{Lada:1992wc,sch-sta,Lada}. 
A closely related cousin of L$_{\infty}$ algebras are A$_{\infty}$ algebras, which generalize 
associative algebras to structures without associativity \cite{stasheff}.

Our goal in this paper is to prove general theorems about the existence of L$_{\infty}$ structures 
for given `initial data' such as an antisymmetric bracket 
and to discuss their possible field theory realizations. 
First, as a warm-up, we answer the following natural question: 
Given a vector space $V$ with an antisymmetric bracket $[\, \cdot,\cdot\,]$, under which conditions 
can this algebra be extended to an L$_{\infty}$ algebra with $\ell_2(v,w)= [v,w]$? 
We will show 
that this is always 
possible.  More specifically, we will prove the following theorem: The graded vector space 
$X=X_1+X_0$, where $X_0=V$ is the space of degree zero and $X_1=V^*$ is isomorphic to $V$ 
and of degree one, carries a 2-term L$_{\infty}$ structure, meaning that the highest non-trivial 
product is $\ell_3$, which encodes the `Jacobiator' (i.e., the anomaly due to 
the failure of the original bracket to 
satisfy the Jacobi identity). 
We have been informed that this theorem is known to some experts, and one instance of it 
has been stated in \cite{Mylonas:2012pg}, but we have not been able to find a proof in the literature. 
(See also \cite{Daily,DailyLada} for examples of
finite-dimensional L$_{\infty}$ algebras).

At first sight the above theorem may shed doubt on the usefulness of L$_{\infty}$ algebras, since 
it states that 
\textit{any} generally non-Lie algebra can be extended to an L$_{\infty}$ algebra. 
It should be emphasized, however, that for a generic bracket 
the resulting 
structure is quite degenerate in that the 2-term L$_{\infty}$ algebra may not be extendable further 
in a non-trivial way, 
say by including a vector space $X_{-1}$. 
Such extensions are particularly important 
for applications in theoretical physics as here $X_{-1}$ encodes 
the `space of physical fields', $X_0$ the space of `gauge parameters' and 
$X_{1}$ the space of `trivial parameters' whose action on fields vanishes \cite{Hohm:2017pnh}. 
Thus, if $X_{1}$ is isomorphic to $X_0$ there is 
no non-trivial action of $X_0$ on the physical fields and hence no genuine 
field theory realization of the L$_{\infty}$ algebra. 
In order to obtain non-trivial field theory realizations 
we will next prove a much more general theorem that covers the case of the Jacobiator 
being of a special form.  
Specifically, we will prove that if the Jacobiator takes values in the image of a linear 
operator that defines an ideal of the original algebra then there exists 
a 3-term L$_{\infty}$ algebra whose highest bracket in general is a non-trivial $\ell_4$. 
A special case is the Courant bracket investigated by Roytenberg and Weinstein \cite{roytenberg-weinstein}, 
for which the 4-bracket trivializes, but which is extendable and realized in string theory, in the 
form of double field theory \cite{Deser:2014mxa,Deser:2016qkw,Hohm:2017pnh}.

We will illustrate these results with examples. 
Our investigation arose in fact out of the question whether the non-associative 
octonions (more precisely, the 7-dimensional 
commutator algebra of imaginary octonions) can be viewed as part of an L$_{\infty}$ algebra. 
Our first theorem implies that the answer is affirmative, with the total graded space being 14-dimensional, 
which we will see is minimal. However, given the theorem, 
the existence of this L$_{\infty}$ structure  does not express a non-trivial fact about the octonions.
Moreover, this L$_{\infty}$ structure is not extendable, which implies with the results of 
\cite{Hohm:2017pnh} that the octonions, at least when 
realized as a 2-term L$_{\infty}$ algebra, 
cannot realize a non-trivial gauge symmetry in field theory.

As recently discovered in  \cite{Gunaydin:2016axc} and further investigated in \cite{Kupriyanov:2017oob,Lust:2017bgx},
the octonions are related to the phase space of non-geometric backgrounds in M-theory (non-geometric R-flux or non-geometric Kaluza-Klein monopoles in M-theory).
Furthermore,
a contraction 
of the octonions leads to the string theory `R-flux algebra'  of \cite{Blumenhagen:2010hj,Lust:2010iy,Blumenhagen:2011ph,Mylonas:2012pg,Bakas:2013jwa}  
and also to the `magnetic monopole algebra' of   \cite{Jackiw:1984rd,Grossman:1984fs,Wu:1984wr,Jackiw:1985hq,Mickelsson:1985fa,Gunaydin:1985ur,Bakas:2013jwa}.
The Jacobiator of the R-flux algebra only takes 
values in a one-dimensional subspace,
and therefore these contracted non-associative algebras may in fact be extendable. 
Here it is sufficient to take $X_1$ to be one-dimensional, leading to an 8-dimensional L$_{\infty}$ algebra.
(A 14-dimensional and hence non-minimal  L$_{\infty}$ realization of the R-flux algebra has already 
been given in \cite{Mylonas:2012pg}.) 

The remainder of this paper is organized as follows. In sec.~2 we briefly review the axioms of 
L$_{\infty}$ algebras. In sec.~3 we prove the theorem that for arbitrary 2-bracket as initial data  
there is an L$_{\infty}$ structure on the `doubled' vector space. 
This theorem will then be significantly generalized in sec.~4. 
In sec.~5 we discuss examples, such as the octonions, the `R-flux algebra', and the 
Courant algebroid. In the appendix we prove an analogous result for A$_{\infty}$ algebras.

\section{Axioms of L$_{\infty}$ algebras}

We begin by stating the axioms of an L$_{\infty}$ algebra. It is defined on a 
graded vector space 
 \be
   X \ = \  \bigoplus_{n \in \mathbb{Z} }  X_n \,, 
 \ee  
and we refer to elements in $X_n$ as having degree $n$. 
We also refer to algebras with $X_n=0$ for all $n$ with $|n|\geq k$ as a $k$-term  L$_{\infty}$ algebra. 
There are a potentially infinite number of generalized multi-linear products or brackets $\ell_k$ having 
$k$ inputs and  
intrinsic degree $k-2$, meaning that they take values in a vector space whose degree 
is given  by 
\be
   \hbox{deg}  \, ( \ell_k (x_1 , \ldots , x_k) ) \ = \   k-2 + \sum_{i=1}^k  \hbox{deg} (x_i) \,. 
\ee 
For instance, $\ell_1$ has intrinsic degree $-1$, implying that it acts on the graded vector space 
according to 
 \be\label{exactsequence}
  \cdots \quad \rightarrow \quad X_1 \quad \xrightarrow{\ell_1} \quad X_0 \quad \xrightarrow{\ell_1} 
  \quad X_{-1}\quad 
  \rightarrow \quad \cdots 
 \ee
Moreover, the brackets are \textit{graded (anti-)commutative} in that, e.g.,  $\ell_2$ satisfies 
 \be\label{gradedCOMM}
  \ell_2(x_1,x_2) \ = \  (-1)^{1+ x_1 x_2}\,  \ell_2(x_2,x_1)\;,  
   \ee
and similarly for all other brackets.

The brackets have to satisfy a (potentially infinite) 
number of generalized Jacobi identities. In order to state these identities we have to define the 
Koszul sign $\epsilon (\sigma;x )$ for any $\sigma$ in the permutation group of $k$ objects 
and a choice $x=(x_1,\ldots, x_k)$ of $k$ such objects. It can be defined implicitly by considering a graded commutative algebra 
with 
 \be
x_i \wedge x_j \ = \ (-1)^{x_i x_j}   \,  x_j \wedge x_i \,,   \quad \forall i, j \,, 
\ee
where in exponents $x_i$ denotes the degree of the corresponding element. 
The Koszul sign is then inferred from 
\be
 x_1\wedge \ldots  \wedge x_k  \ = \ \epsilon (\sigma; x)  \   x_{\sigma(1)} \wedge \ldots   \wedge \, x_{\sigma(k)} \,. 
\ee
The L$_{\infty}$ relations are given by 
\be
\label{main-Linty-identity}
\sum_{i+j= n+1}  (-1)^{i(j-1)} \sum_\sigma (-1)^\sigma \epsilon (\sigma; x) \, \ell_j  \bigl(  \ell_i ( x_{\sigma(1)}  \,, \, \ldots\,, x_{\sigma(i)} ) \,, \, x_{\sigma(i+1)}, \, \ldots \, x_{\sigma (n)} \bigr) \ = \ 0\,, 
\ee
for each $n=1,2,3,\ldots $, which indicates the total number of inputs.  
Here $(-1)^\sigma$ gives a plus sign if the permutation is even and a minus sign if the permutation is odd.
Moreover, the inner sum runs, for a given $i, j\geq 1$, over all permutations $\sigma$ of $n$ objects 
whose arguments are partially ordered (`unshuffles'), satisfying 
\be
\sigma(1) \leq  \, \cdots \, \leq  \, \sigma(i) \,,  \qquad 
\sigma(i+1) \leq  \, \cdots \, \leq  \, \sigma(n) \,.
\ee

We will now state these relations explicitly for the values of $n$ relevant for our subsequent analysis. 
For $n=1$  the identity reduces to 
 \be\label{l1-identity}
  \ell_1 ( \ell_1 (x)) \ = \ 0 \,, 
  \ee
stating that $\ell_1$ is nilpotent, so that (\ref{exactsequence}) is a chain complex. 
For $n=2$ the identity reads 
  \be
  \label{L2L1}
 \ell_1(\ell_2(x_1,x_2)) \ = \ \ell_2(\ell_1(x_1),x_2) + (-1)^{x_1}\ell_2(x_1,\ell_1(x_2))\;, 
 \ee
meaning that $\ell_1$ acts like a derivation on the product $\ell_2$. 
For $n=3$ one obtains 
 \be\label{L3L1}
  \begin{split}
  0    \  &=  \ \ell_1(\ell_3 (x_1,x_2, x_3))  \\
&   + \ell_3(\ell_1 (x_1) ,x_2, x_3) 
  +  (-1)^{x_1}  \ell_3( x_1 ,\ell_1(x_2), x_3)
    +  (-1)^{x_1+ x_2}  \ell_3( x_1 ,x_2, \ell_1(x_3)) \\
 & + \ell_2(\ell_2(x_1,x_2),x_3) + (-1)^{(x_1+ x_2) x_3}\ell_2(\ell_2(x_3,x_1),x_2) 
   +(-1)^{(x_2+ x_3) x_1 }\ell_2(\ell_2(x_2,x_3),x_1) \;.  
  \end{split} 
 \ee
We recognize the last line as the usual Jacobiator.  
Thus, this relation encodes the failure of the 2-bracket to satisfy the Jacobi identity in terms 
of a 1- and 3-bracket and the failure of $\ell_1$ to act as a derivation on $\ell_3$.  
Finally, the  $n=4$ relations read 
\be\label{ell4ell1SIMP}
\begin{split}
{\cal O}(x_1,\ldots,x_4) \ \equiv \ & - \ \ell_2 ( \, \ell_3 (x_1, x_2, x_3) , x_4)  
\ + \ (-1)^{x_3 x_4}\, \ell_2  ( \, \ell_3 (x_1, x_2, x_4) , x_3) 
 \\ 
 & + \ (-1)^{(1+x_1)x_2} \ell_2 (x_2, \ell_3 (x_1, x_3, x_4)) 
 \ - \ (-1)^{x_1} \ell_2 (x_1, \ell_3 (x_2, x_3, x_4) ) \\
 & + \ \ell_3 ( \ell_2 (x_1, x_2 ) , x_3, x_4)   
 \ + \ (-1)^{1 + x_2 x_3} \   \ell_3 ( \ell_2 (x_1, x_3 ) , x_2, x_4) \\
 &  + \ (-1)^{x_4 (x_2 + x_3)} \ell_3 ( \ell_2 (x_1, x_4 ) , x_2, x_3)   
 \ - \ \ell_3 ( x_1, \ell_2 (x_2, x_3 ) ,  x_4)  \\
 &  + \ (-1)^{x_3x_4}\ell_3 (x_1,  \ell_2 (x_2, x_4 ) , x_3)  
 \ + \ \ell_3 (x_1,  x_2, \ell_2 (x_3, x_4 ) )   \\
 \ = \ & -\ell_1 ( \, \ell_4 ( x_1, x_2, x_3, x_4)) \\
&+ \ \ell_4  (\ell_1 (x_1), x_2, x_3, x_4) 
 \ + \ (-1)^{x_1}  \ell_4 ( x_1, \ell_1 (x_2), x_3, x_4)   \\
 &  + \ (-1)^{x_1+x_2}  \ell_4 ( x_1, x_2, \ell_1 (x_3), x_4)
  \ + \  (-1)^{x_1+ x_2 + x_3}  \ell_4 ( x_1, x_2, x_3, \ell_1 (x_4)) \,, 
 \end{split}
\ee
where we named the l.h.s.~${\cal O}(x_1,\ldots,x_4)$ for later convenience. 
For a 2-term L$_{\infty}$ algebra there are no 4-brackets and hence
the above right-hand side is zero. The $n=4$ relation then poses a non-trivial constraint on $\ell_2$
and $\ell_3$, while all higher L$_{\infty}$ relations will be automatically satisfied.

\section{A warm-up theorem}\label{FirstTheorem}

We now prove the first theorem stated in the introduction. Consider 
an algebra $(V,[\, \cdot,\cdot\,])$ with  bilinear antisymmetric 2-bracket, i.e.
\begin{equation}
[v,w] \ = \ -[w,v]\qquad \forall v,w\, \in \, V\,,  \label{4.1}
\end{equation}
but we do \textit{not} assume that the bracket satisfies any further constraints.  
In particular, the Jacobi identity is generally not satisfied, so that the \textit{Jacobiator} 
\begin{equation}
{\rm Jac }(u,v,w) \ \equiv \ [[u,v],w]+[[v,w],u]+[[w,u],v]\;,   \label{4.2}
\end{equation}
in general is non-zero. 
We then have the following \\[1ex]
\textit{Theorem 1:}\\
The graded vector space 
 \be\label{Xgrading}
  X \ = \  X_1  +  X_{0} \;, 
 \ee
where $X_0=V$ and $X_1=V^*$ with $V^*$ a vector space isomorphic to $V$, carries a 
2-term L$_{\infty}$ structure whose non-trivial brackets are given by 
\begin{align}
\ell_1(v^*)\, &=\, v\;, \\
\ell_2(v,w)\, &=\, [v,w]\;, \\
\ell_2(v^*,w)\,&=\, [v, w]^*\label{4.5}\;, \\
\ell_3(u,v,w)\,&=\, - {\rm Jac}(u,v,w)^*\label{4.7} \;. 
\end{align}
\textit{Comment:}\\
We denote the elements of $V^*$ by $v^*, w^*$, etc., and the isomorphism by 
 \be\label{isomorphism}
  \phantom{f}^*:V\; \rightarrow\; V^*\,,\qquad  v\, \mapsto \, v^*\;, 
 \ee 
and similarly for its inverse. For instance, if $V$ carries a non-degenerate metric we can take 
$V^*$ to be the dual vector space of $V$ and the isomorphism to be the canonical isomorphism. 
(More simply, we can think of $V^*$ as a second copy of $V$ and of the isomorphism as the identity, 
but at least for notational reasons it is important to view $V$ and $V^*$ as different objects.) 
\\[0.5ex]
\textit{Proof:}\\
The proof proceeds straightforwardly by fixing the products so that the $n=1, 2, 3$ 
relations are partially satisfied and then verifying that in fact all L$_{\infty}$ relations are satisfied. 
First, $\ell_1$ maps $X_1=V^*$ to $X_0=V$, and we take it to be given by the (inverse) isomorphism 
(\ref{isomorphism}),  
 \be \label{ell1action}
 \forall   v^*\in X_1, v\in X_0:\qquad  \ell_1(v^*) \ = \ v\;, \qquad \ell_1(v)\ = \ 0 \;. 
 \ee  
The second relation in here is necessary because  there is no space 
$X_{-1}$ in (\ref{Xgrading}). The $n=1$ relations $\ell_1^2=0$ then hold trivially.   

Next, we fix the $\ell_2$ product by requiring $\ell_2(v,w)=[v,w]$ on $X_0=V$ and imposing 
the $n=2$ relation (\ref{L2L1}). For arguments of total degree 0 this relation is trivial because 
of the second relation in (\ref{ell1action}). 
For arguments of total degree 1 we have  
\begin{align}
 \ell_1(\ell_2(v^*,w))\,= \, \ell_2(\ell_1(v^*),w) - \ell_2(v^*, \ell_1(w)) \ = \ \ell_2(v,w) \ = \ [v,w]\;, 
\end{align}
where we used (\ref{ell1action}). 
Using (\ref{ell1action}) on the l.h.s.~we infer 
 \be\label{ell2vwstar}
   \ell_2(v^*, w) \ = \   [v,w]^* \qquad \Leftrightarrow \qquad 
   \ell_2(w, v^*) \ = \ [w,v]^*\;. 
 \ee  
Since there is no space $X_{2}$ we have $\ell_2(v^*,w^*)=0$.  This is consistent with the 
$n=2$ relation (\ref{L2L1}) for arguments of total degree 2: 
\begin{align}
0 \ = \ \ell_1(\ell_2(v^*,w^*)) \ = \ \ell_2(v,w^*)\, -\, \ell_2(v^*,w) \ = \ [v,w]^*\,-\, [v,w]^*\, =\, 0\;, 
\end{align}
where we used (\ref{ell2vwstar}). Thus, all $n=2$ relations are satisfied. 

Let us now consider the $n=3$ relations (\ref{L3L1}). For arguments 
of total degree 0 (i.e., all taking values in $X_0$), it reads  
\begin{align}
0 \ &= \ \ell_1(\ell_3(u,v,w))+\ell_2(\ell_2(u,v),w)+\ell_2(\ell_2(v,w),u)+\ell_2(\ell_2(w,u),v)\\ \nonumber 
 \ &= \ \ell_1(\ell_3(u,v,w)) +{\rm Jac} (u,v,w)\;, 
\end{align}
from which we infer 
 \be\label{ell3assignment}
  \ell_3(u,v,w) \ = \ -{\rm Jac} (u,v,w)^* \ \in \ X_1\;. 
 \ee
Due to the antisymmetry of the bracket $[\cdot\,,\cdot ]$, the Jacobiator is 
completely antisymmetric in all arguments,  and (\ref{ell3assignment}) is consistent 
with the required graded commutativity of $\ell_3$. Since there is no space $X_2$, 
$\ell_3$ is trivial for any arguments in $X_1$.  We have thus determined all non-trivial 
$n$-brackets.

So far we have verified the $n=1,2$ relations and the $n=3$ relation for arguments 
of total degree 0. We now verify the remaining L$_{\infty}$ relations. 
The $n=3$ relation for arguments of total degree 1 reads:
\begin{align}
0 \ &= \ \ell_3(\ell_1(u^*),v,w)+\ell_2(\ell_2(u^*,v),w)+\ell_2(\ell_2(w,u^*),v)+\ell_2(\ell_2(v,w),u^*)\\ \nonumber 
\ &= \ - {\rm Jac }(u,v,w)^*+{\rm Jac} (u,v,w)^*\;, 
\end{align}
and is thus satisfied. 
The $n=3$ relations for arguments of total degree larger than 1 are trivially satisfied, completing
 the proof of all $n=3$ relations. 

Finally, we have to verify the $n=4$ relations. Since there is no non-trivial $\ell_4$ these 
require that the left-hand side of (\ref{ell4ell1SIMP}) vanishes identically for $\ell_2$ and $\ell_3$
defined above. This follows by a direct computation that we display in detail. 
First, for arguments $v_1, v_2, v_3, v_4\in X_0$ of total degree 0 one may verify that 
(\ref{ell4ell1SIMP}) is completely antisymmetric in the four arguments. Writing $\sum_{\rm anti}$
for the totally antisymmetrized sum (carrying $4!=24$ terms and pre-factor $\frac{1}{4!}$) we then 
compute for the left-hand side of (\ref{ell4ell1SIMP})
 \be\label{n=4firsttheorem}
  \begin{split}
   {\cal O}(v_1,\ldots, v_4) \ &= \ \sum_{\rm anti}\Big(-4\, \ell_2(\ell_3(v_1,v_2,v_3), v_4)
   +6\, \ell_3(\ell_2(v_1,v_2), v_3, v_4)\, \Big)\\
     \ &= \  \sum_{\rm anti}\Big(4\, [{\rm Jac}(v_1,v_2,v_3), v_4]^*
   -6\, {\rm Jac}([v_1,v_2], v_3, v_4)^*\, \Big) \\
    \ &= \  \sum_{\rm anti}\Big(12\, [[[v_1,v_2],v_3], v_4]^*
   -6\, \big(2\,[[[v_1,v_2], v_3], v_4]^* +[[v_3,v_4],[v_1,v_2]]^*\big)\, \Big)\\
    \ &= \ 0\;. 
  \end{split}
 \ee   
Here we used repeatedly the total antisymmetry in the four arguments, 
in particular in the last step that under the sum $[[v_3,v_4],[v_1,v_2]]^*$ then vanishes. 
The $n=4$ relations for arguments of total degree 1 or higher 
are trivially satisfied because they would have to take values in spaces of degree $2$ or higher,
which do not exist. The L$_{\infty}$ relations for $n>4$ are trivially satisfied for the same reason.
This completes the proof. $\square$

\section{Main theorem}

The above theorem states that an arbitrary bracket can be extended to an L$_{\infty}$ algebra. 
For generic brackets, this L$_{\infty}$ structure is, however, quite degenerate in that it may not 
be extendable further, say by adding a further space $X_{-1}$. 
Indeed, if the violation of the Jacobi identity is `maximal' and the Jacobiator 
takes values in all of $V$, the space $X_{1}$ has to be as large as $V$, and 
the image of the map $\ell_1: X_1\rightarrow X_0$ equals $X_0=V$.  
Consequently,  one cannot introduce a further space $X_{-1}$ together with a non-trivial 
$\ell_1: X_0\rightarrow X_{-1}$ satisfying $\ell_1^2=0$. 
Since in physical applications $X_{-1}$ serves as the space of fields, such brackets do not lead to 
L$_{\infty}$ algebras encoding  a non-trivial gauge symmetry. 

More interesting situations arise when the Jacobiator takes values in a proper subspace $U\subset V$, 
for then it is sufficient to set $X_1=U$  and to take $\ell_1=\iota$ to be the `inclusion' defined for any $u\in U$ 
by $\iota(u)=u$, viewing $u$ as an element of $V$. Indeed, it is easy to verify, 
provided the subspace forms an ideal (i.e., $\forall u\in U, v\in V: [u,v]\in U$), that the above proof goes through as before. 
In this case, further extensions of the L$_{\infty}$ algebra may exist.
In the following we will prove a yet more general theorem that is applicable to situations where 
the Jacobiator takes values in the image of a linear map that itself may have a non-trivial kernel. 
Then there is an extension to a 3-term L$_{\infty}$ algebra that generally requires a non-trivial 
4-bracket: 
\\[2ex]
\textit{Theorem 2:}\\[0.5ex] 
Let $(V,[\, \cdot,\cdot\,])$ be an algebra with bilinear antisymmetric 2-bracket as in sec.~\ref{FirstTheorem}, 
and let ${\cal D}:\, U\rightarrow V$
be a linear map 
satisfying the closure conditions  
  \be\label{bracketCOMP}
  [{\rm Im}({\cal D}), V] \ \subset \ {\rm Im}({\cal D})\;,
  \ee 
together with the Jacobiator relation 
 \be\label{JACisMAP}
  \forall v_1, v_2, v_3 \in V : \; \; {\rm Jac}(v_1,v_2,v_3) \ \in \ {\rm Im}({\cal D}) \;, 
 \ee
where ${\rm Im}({\cal D})$ and ${\rm Ker}({\cal D})$ denote image and kernel of  ${\cal D}$, respectively. 
Then there exists a 3-term L$_{\infty}$ structure with $\ell_2(v,w)=[v,w]$ 
on the graded vector space  
with 
 \be\label{longerchain}
  X_2 \;\xrightarrow{\ell_1=\iota } \; X_1 \; \xrightarrow{\ell_1={\cal D}} 
  \; X_{0}\;, 
 \ee
where $X_0=V$, $X_1=U$, $X_2={\rm Ker}({\cal D})$ and $\iota$ denotes the 
inclusion of ${\rm Ker}({\cal D})$ into $U$. The highest non-trivial bracket in general is given by the 
4-bracket (and the complete list of non-trivial brackets is 
given in eq.~(\ref{ALLproducts}) below). 
\\[2ex]
\noindent 
\textit{Notation and comments:}\\[0.5ex]
We denote the elements of $V$ by $v, w, \ldots$, the elements of $U$ by $\alpha,\beta, \ldots$
and the elements of ${\rm Ker}({\cal D})$ by $c, c', \ldots$
The condition (\ref{JACisMAP}) implies that there is a multi-linear and totally antisymmetric 
map $f : V^{\otimes 3} \rightarrow U$ so that 
 \be\label{JacISf3}
  \forall v_1, v_2, v_3 \in V : \; \; {\rm Jac}(v_1,v_2,v_3) \ = \ {\cal D}f(v_1,v_2,v_3)\;. 
 \ee
The condition (\ref{bracketCOMP}) states that the bracket of an arbitrary $v\in V$ with ${\cal D}\alpha$, 
$\alpha\in U$, lies in the image of ${\cal D}$, i.e., we can write 
 \be\label{closureEXP}
  \forall v\in V, \alpha\in U:\quad [{\cal D}\alpha, v] \ = \ {\cal D}(v(\alpha))\;, \qquad v(\alpha) \ \in \  U\;. 
 \ee
We can think of the operation on the r.h.s.~as defining for each $v\in V$ a map on $U$,  
$\alpha\mapsto v(\alpha)\in U$. 
This map is defined by (\ref{closureEXP}) only up contributions in the kernel, as is the function $f$ 
in (\ref{JacISf3}), 
but the following 
construction goes through for any choice of functions 
satisfying (\ref{closureEXP}), (\ref{JacISf3}).\footnote{The algebras resulting for different choices 
of theses functions are almost certainly equivalent under suitably defined L$_{\infty}$ isomorphisms, see, e.g.,  \cite{MARKLmonograph}, but we leave a detailed analysis for future work.} 
\\[0.5ex]
\textit{ Proof:}\\
As for Theorem 1, the proof proceeds by determining the $n$-brackets from the L$_{\infty}$ relations 
as far as possible and then proving that in fact all relations are satisfied. 
The $n=1$ relations $\ell_1^2=0$ for $\ell_1$ defined in (\ref{longerchain}) are satisfied by definition 
since ${\cal D}(\iota(c))=0$ for all $c \in {\rm Ker}({\cal D})$.  
In the following we systematically go through all relations for $n=1,\ldots, 5$. 

 \textit{$n=2$ relations:} 
 The $n=2$ relations are satisfied for arguments of total degree zero, since $\ell_1$ acts trivially on 
$X_{0}$. For arguments $\alpha\in X_1$, $v\in X_0$ of total degree 1 we need 
 \be
  \ell_1(\ell_2(\alpha, v)) \ = \ \ell_2(\ell_1(\alpha), v) \ = \ [{\cal D}\alpha, v] \ = \ {\cal D}(v(\alpha))\;, 
 \ee
where we used (\ref{closureEXP}). As the l.h.s~equals ${\cal D}(\ell_2(\alpha, v))$, this relation 
is satisfied if we set 
 \be\label{ell2vandalpha}
  \ell_2(\alpha, v) \ = \ v(\alpha) \ \in \ X_1\;. 
 \ee
For arguments $\alpha,\beta\in X_1$ of total weight 2 we compute  
 \be
 \begin{split}
  \ell_1(\ell_2(\alpha,\beta)) \ &= \ \ell_2(\ell_1(\alpha), \beta) -  \ell_2(\alpha, \ell_1(\beta)) \ = \ 
  -\ell_2(\beta, {\cal D}\alpha) -\ell_2(\alpha, {\cal D}\beta)\\
  \ &= \ -({\cal D}\alpha)(\beta)-({\cal D}\beta)(\alpha)\;, 
 \end{split} 
 \ee
using (\ref{ell2vandalpha}) in the last step.  
As $\ell_1$ on the l.h.s~acts by inclusion, we can satisfy this relation by setting 
 \be\label{ell2alphabeta}
  \ell_2(\alpha,\beta) \ = \  -({\cal D}\alpha)(\beta)-({\cal D}\beta)(\alpha) \ \in \ {\rm Ker}({\cal D})\;, 
 \ee
but it remains to prove that the r.h.s.~indeed takes values in the kernel. This follows  
 by setting $v={\cal D}\beta$ in (\ref{closureEXP}): 
  \be
    [{\cal D}\alpha, {\cal D}\beta] \ = \ {\cal D}(({\cal D}\beta)(\alpha)) \quad \Rightarrow \quad 
 {\cal D}\big(({\cal D}\alpha)(\beta)+({\cal D}\beta)(\alpha)\big) \ = \ 0 \;, 
 \ee 
using that the bracket is antisymmetric. 
Note that 
(\ref{ell2alphabeta}) is properly symmetric in its two arguments, in agreement with the graded commutativity 
(\ref{gradedCOMM}). Another choice of arguments of total degree 2 is $v\in X_0, c \in X_2$, 
for which we require 
 \be
  \ell_1(\ell_2(v,c)) \ = \ \ell_2(\ell_1(v), c) + \ell_2(v, \ell_1(c)) \ = \ \ell_2(v,\iota(c)) 
  \ = \ -v(\iota(c)) \;, 
 \ee
where we used (\ref{ell2vandalpha}) in the last step, recalling $\iota(c)\in X_1$. 
Thus, using $\ell_1=\iota$ on the l.h.s.~together with the graded symmetry we have 
 \be
  \iota(\ell_2(c,v)) \ = \  v(\iota(c))\;. 
 \ee
We can also write this as\footnote{Here we employ
the map on $X_2$ induced by $v(\alpha)$ via $v(c):=v(\iota(c))$, which lies in 
${\rm Ker}({\cal D})$ as a consequence of ${\cal D}c=0$ and (\ref{closureEXP})}
 \be\label{cvBRACKET}
  \forall c\in X_2, v\in X_0:\quad \ell_2(c, v) \ = \ v(c) \ \in  \ X_2\;. 
 \ee
We next consider arguments $c\in X_2, \alpha\in X_1$ of total degree 3, for 
which $\ell_2$ must vanish as there is 
no vector space $X_3$. This leads to a constraint from the $n=2$ relation: 
 \be
  0 \ = \ \ell_1(\ell_2(c,\alpha)) \ = \ \ell_2(\iota(c), \alpha) + \ell_2(c, {\cal D}\alpha) 
  \ = \  -({\cal D}\alpha)(c)+ \ell_2(c, {\cal D}\alpha) \;, 
 \ee 
where we used (\ref{ell2alphabeta}) and ${\cal D}c=0$. This relation is satisfied for (\ref{cvBRACKET}). 
Finally, the $n=2$ relations are trivially satisfied for arguments of total degree 4 or higher, 
completing the proof of all $n=2$ relations. 

\textit{$n=3$ relations:} 
We now consider the $n=3$ relations for arguments $v_1, v_2, v_3\in X_0$ of total degree zero: 
 \be
  0 \ = \ \ell_1(\ell_3(v_1, v_2, v_3)) + {\rm Jac}(v_1,v_2,v_3)\;. 
 \ee
Recalling (\ref{JacISf3}) and that $\ell_1={\cal D}$ when acting on $X_1$, we infer that this 
relation is satisfied for 
 \be\label{ell3JAC}
  \ell_3(v_1,v_2,v_3) \ = \ -f(v_1,v_2,v_3) \ \in \ X_1\;. 
 \ee
Next, for arguments $\alpha\in X_1, v_1, v_2\in X_0$ of total weight 1 the $n=3$ relation reads 
 \be\label{n=3onemorestep}
 \begin{split}
  0 \ = \ &\,\ell_1(\ell_3(\alpha, v_1, v_2)) + \ell_3(\ell_1(\alpha), v_1, v_2) \\
  &\, +\ell_2(\ell_2(\alpha, v_1), v_2) + \ell_2(\ell_2(v_2, \alpha), v_1) + \ell_2(\ell_2(v_1, v_2),\alpha)\\
  \ = \ &\, \iota(\ell_3(\alpha, v_1,v_2)) - f({\cal D}\alpha, v_1, v_2) + v_2(v_1(\alpha))-v_1(v_2(\alpha))
  -[v_1, v_2](\alpha)\;, 
 \end{split}
 \ee      
where we used repeatedly (\ref{ell2vandalpha}). Moreover, we used  (\ref{ell3JAC}) 
and that $\ell_3(\alpha, v_1, v_2)\in X_2$ on which $\ell_1$ acts as the inclusion. 
We will next prove that the function 
 \be\label{offdiagonalell30}
  g(\alpha, v_1,v_2) \ \equiv \ f({\cal D}\alpha, v_1, v_2) 
  + [v_1, v_2](\alpha) + v_1(v_2(\alpha)) - v_2(v_1(\alpha))\;, 
 \ee
takes values in the subspace ${\rm Ker}({\cal D})$.  
We have to prove that the r.h.s.~is annihilated by ${\cal D}$. 
To this end we compute for the first term with (\ref{JacISf3}) 
 \be
 \begin{split}
  {\cal D}f({\cal D}\alpha, v_1, v_2) \ &= \ {\rm Jac}({\cal D}\alpha, v_1, v_2) \\
   \ &= \ 
  [[{\cal D}\alpha,v_1],v_2]+[[v_2, {\cal D}\alpha], v_1] + [[v_1,v_2],{\cal D}\alpha]\\
   \ &= \ [{\cal D}(v_1(\alpha)), v_2]-[{\cal D}(v_2(\alpha)), v_1] - {\cal D}([v_1, v_2](\alpha))  \\
   \ &= \ {\cal D}\Big(v_2(v_1(\alpha)) - v_1(v_2(\alpha)) - [v_1, v_2](\alpha)\Big)\;, 
 \end{split}
 \ee 
where we repeatedly used (\ref{closureEXP}). This show that the r.h.s.~of (\ref{offdiagonalell30}) 
is annihilated by ${\cal D}$, proving that $g$ takes values in $X_2={\rm Ker}({\cal D})$. 
We can thus satisfy (\ref{n=3onemorestep}) by setting  
 \be\label{offdiagonalell3}
  \ell_3(\alpha, v_1,v_2) \ = \ g(\alpha, v_1,v_2) \ \in X_2 \;. 
 \ee

We next recall that there can be no non-trivial $\ell_3$ for arguments $\alpha_1, \alpha_2\in X_1, v\in X_0$ 
of total degree 2. Thus, the $n=3$ relation for these arguments has to be satisfied for the 
products already defined. We then compute from (\ref{L3L1}), noting that it is symmetric in $\alpha_1, \alpha_2$
and writing $\sum_{\rm sym}$ for the symmetrized sum,  
 \be
  \begin{split}
   0 \ = \ \sum_{\rm sym} \big(& 2\, \ell_3(\ell_1(\alpha_1), \alpha_2, v) + \ell_2(\ell_2(\alpha_1, \alpha_2), v)
   +2\, \ell_2(\ell_2(v, \alpha_1), \alpha_2)\big)\\
   \ = \ \sum_{\rm sym}\big(& -2\, \ell_3(\alpha_1, {\cal D} \alpha_2, v)
   -2\, \ell_2(({\cal D}\alpha_1)(\alpha_2), v)-2\,\ell_2(v(\alpha_1), \alpha_2)\big)\\
   \ = \ \sum_{\rm sym}\big(&-2\, f({\cal D}\alpha_1, {\cal D}\alpha_2, v)-2\, [{\cal D}\alpha_2, v](\alpha_1)
   -2\, ({\cal D}\alpha_2)(v(\alpha_1))  + 2\, v(({\cal D}\alpha_2)(\alpha_1)) \\[-1.5ex]
   &-2\, v(({\cal D}\alpha_1)(\alpha_2)) +2\, ({\cal D}(v(\alpha_1)))(\alpha_2) +2\, ({\cal D}\alpha_2)(v(\alpha_1))\big)\;,  
  \end{split}
 \ee  
where we used (\ref{ell2vandalpha}), (\ref{ell2alphabeta}) and, in the third equality, (\ref{offdiagonalell3}). 
It is now easy to see that under the symmetrized sum all terms cancel, using in particular 
that $f$ is totally antisymmetric. Thus, this $n=3$ relation is satisfied. 
Since the $n=3$ relations for total degree 3 or higher are trivially satisfied, we have completed the proof 
of all $n=3$ relations. 

\textit{$n=4$ relations:} We consider the $n=4$ relations (\ref{ell4ell1SIMP}) for arguments of total 
degree 0. Precisely as in (\ref{n=4firsttheorem}) we compute 
 \be\label{n=4BigStep}
   \begin{split}
   {\cal O}(v_1,\ldots, v_4) \ &= \ \sum_{\rm anti}\Big(-4\, \ell_2(\ell_3(v_1,v_2,v_3), v_4)
   +6\, \ell_3(\ell_2(v_1,v_2), v_3, v_4)\, \Big)\\
    \ &= \ \sum_{\rm anti}\Big(-4\, v_1(f(v_2, v_3, v_4)) - 6\, f([v_1, v_2], v_3, v_4)\Big)\;. 
  \end{split}
 \ee  
In contrast to (\ref{n=4firsttheorem}) this is not zero in general, but we can now have 
a non-trivial $\ell_4$ taking values in $X_2$.
We next prove that the function defined by 
 \be\label{DEFh}
  h(v_1,\ldots, v_4) \ \equiv \ \sum_{\rm anti}\Big(4\, v_1(f(v_2, v_3, v_4)) + 6\, f([v_1, v_2], v_3, v_4)\Big)\;, 
 \ee
takes values in ${\rm Ker}({\cal D})$. To this end we have to show that it is annihilated by ${\cal D}$: 
 \be 
 \begin{split}
  {\cal D}(h(v_1,\ldots, v_4)) \ &= \ 
  \sum_{\rm anti}\Big(4\, {\cal D}(v_1(f(v_2, v_3, v_4))) + 6\, {\cal D}f([v_1, v_2], v_3, v_4)\Big) \\
  \ &= \ 
  \sum_{\rm anti}\Big(4\, [{\rm Jac}(v_2, v_3, v_4),v_1] + 6\, {\rm Jac}([v_1, v_2], v_3, v_4)\Big)\;.  \\
  \ &= \ 0\;. 
 \end{split} 
 \ee 
 Thus, the $n=4$ relation can be satisfied 
by setting 
 \be\label{ell4fourv}
  \ell_4(v_1,\ldots, v_4) \ = \  h(v_1,\ldots, v_4) \ \in \ X_2\;.  
 \ee

We have now determined all non-trivial brackets, which we summarize here: 
 \be \label{ALLproducts}
  \begin{split}
   c\in X_2:&\quad \ell_1(c) \ = \ \iota(c) \ = \ c  \ \in \ X_1\;, \\
   \alpha\in X_1:&\quad \ell_1(\alpha) \ = \ {\cal D}\alpha \ \in \ X_0\;, \\
   v, w\in X_0:& \quad \ell_2(v,w) \ = \ [v,w] \ \in \ X_0\;, \\
   \alpha\in X_1, v\in X_0: &\quad \ell_2(\alpha, v) \ = \ v(\alpha) \ \in \ X_1\;, \\
   c\in X_2, v\in X_0: &\quad \ell_2(c, v) \ = \ v(c) \ \in  \ X_2\;, \\
   \alpha,\beta \in X_1: &\quad 
   \ell_2(\alpha,\beta) \ = \  -({\cal D}\alpha)(\beta)-({\cal D}\beta)(\alpha) \ \in \ X_2\;, \\
   v_1,v_2,v_3\in X_0:& \quad 
    \ell_3(v_1,v_2,v_3) \ = \ -f(v_1,v_2,v_3) \ \in \ X_1\;, \\
    \alpha\in X_1, v_1, v_2\in X_0: &\quad
   \ell_3(\alpha, v_1,v_2) \ = \ 
    g(\alpha, v_1,v_2) \  \in \  X_2\,, \\
  v_1,\ldots, v_4 \in X_0: & \quad 
  \ell_4(v_1,\ldots, v_4)  \ = \   h(v_1,\ldots, v_4) \ \in \ X_2\;, 
  \end{split}
 \ee  
with the functions $g, h$ defined in (\ref{offdiagonalell30}) and (\ref{DEFh}), respectively. 
All further L$_{\infty}$ relations have to be satisfied identically.   
Let us next consider the $n=4$ relations (\ref{ell4ell1SIMP}) for arguments $v_1, v_2, v_3\in X_0, \alpha\in X_1$
of total degree 1. It is easy to see that (\ref{ell4ell1SIMP}) is then totally antisymmetric in $v_1, v_2, v_3$, 
and writing $\sum_{\rm anti}$ for the antisymmetric sum over these three arguments we compute 
 \be\label{BIGELL4STEP}
 \begin{split}
  {\cal O}&(v_1, v_2, v_3,\alpha) \ = \ \sum_{\rm anti}\Big(-\ell_2(\ell_3(v_1, v_2, v_3),\alpha)
  +3\, \ell_2(\ell_3(\alpha, v_1, v_2), v_3) \\[-1.5ex]
  &\qquad\qquad\qquad \qquad\qquad
   +3\, \ell_3(\ell_2(v_1, v_2), v_3, \alpha) + 3\, \ell_3(v_1, v_2, \ell_2(v_3, \alpha)) \Big)\\
  \ &= \ \sum_{\rm anti}\Big(\ell_2(f(v_1, v_2, v_3),\alpha) +3\, v_3(\ell_3(\alpha, v_1, v_2)) 
  +3\, \ell_3([v_1, v_2],v_3,\alpha) - 3\, \ell_3(v_1, v_2, v_3(\alpha))\Big)\\
  \ &= \ \sum_{\rm anti}\Big(-({\cal D}f(v_1, v_2, v_3))(\alpha)-({\cal D}\alpha)(f(v_1,v_2, v_3))\\[-1.5ex]
  &\qquad\qquad +3\, v_3\big(f({\cal D}\alpha, v_1, v_2)+[v_1, v_2](\alpha) +2\, v_1(v_2(\alpha))\big)\\
  &\qquad\qquad+3\big( f({\cal D}\alpha,[v_1, v_2], v_3)+ [[v_1,v_2],v_3](\alpha)
  +[v_1, v_2](v_3(\alpha))- v_3([v_1, v_2](\alpha))\big)\\
  &\qquad\qquad-3\big(f({\cal D}(v_3(\alpha)),v_1, v_2) + [v_1, v_2](v_3(\alpha))+2\, v_1(v_2(v_3(\alpha)))\big)
  \Big)\\
  \ &= \ \sum_{\rm anti}\Big(-({\cal D}\alpha)(f(v_1,v_2, v_3))+3\,v_3(f({\cal D}\alpha, v_1, v_2))\\[-1.5ex]
  &\qquad\qquad
  +3\, f({\cal D}\alpha,[v_1, v_2], v_3)-3\, f([{\cal D}\alpha,v_3],v_1, v_2) \Big)\;, 
 \end{split}
 \ee 
where we used the products already defined, in particular (\ref{offdiagonalell3}), 
and the relation (\ref{JacISf3}) for the Jacobiator. We observe that various terms cancelled 
under the totally antisymmetric sum. In order to satisfy the $n=4$ relation (\ref{ell4ell1SIMP}), 
the remaining terms need to be equal 
to $\ell_4(v_1,v_2,v_3,{\cal D}\alpha)$. To see this note that writing (\ref{ell4fourv}) with an  
antisymmetrized sum over only the first three arguments one obtains  
 \be 
  \begin{split}
   \ell_4(v_1,\ldots, v_4) \ =  \ \sum_{{\rm anti}[v_1,v_2,v_3]}\Big(&\,3\, v_1(f(v_2,v_3,v_4))
    -v_4(f(v_1, v_2,v_3))\\[-2ex]
   &\,+3\, f([v_1, v_2],v_3,v_4) -3\, f([v_4,v_1],v_2,v_3)\Big)\;. 
 \end{split} 
 \ee
Specializing this to $\ell_4(v_1,v_2,v_3,{\cal D}\alpha)$ we infer that it equals (\ref{BIGELL4STEP}), 
completing the proof of this $n=4$ relation. It is easy to see that for arguments of total degree 2 or higher 
the $n=4$ relations are trivially satisfied. Thus, we have verified all $n=4$ relations.

\textit{$n=5$ relations:} We have not displayed the L$_{\infty}$ relations in sec.~2 for $n\geq 5$ 
explicitly as these get increasingly laborious. However, it is easy to see that here the only non-trivial 
$n=5$ relation has arguments $v_1, \ldots, v_5\in X_0$, which are of even degree so that the 
Koszul sign becomes $\epsilon(\sigma;v)=1$. Moreover, $\ell_5$ is trivial, and it is then easy to verify 
that (\ref{main-Linty-identity}) reduces to 
 \be
  \sum_{\rm anti} \Big(10\, \ell_4(\ell_2(v_1, v_2), v_3, v_4, v_5) + 5\,\ell_2(\ell_4(v_1,v_2,v_3,v_4),v_5)
  +10\, \ell_3(\ell_3(v_1, v_2, v_3),v_4, v_5)\Big) \ = \ 0\,,
 \ee 
where the sum antisymmetrizes over all five arguments. Upon inserting the products in (\ref{ALLproducts}), 
it is a straightforward direct calculation, largely analogous to (\ref{BIGELL4STEP}), to verify that 
this relation is identically satisfied. As these are the only non-trivial L$_{\infty}$  relations for $n=5$
or higher, this completes the proof. $\square$ \\[4ex]
\textit{Specializations:}\\[0.5ex]
As a special case of Theorem 2 let us assume that the Jacobiator takes values in a subspace $U\subset V$, 
which forms an ideal of the bracket. In this case we can take ${\cal D}=\iota$ to be the inclusion map 
$U\rightarrow V$. Since its kernel is trivial, we have $X_2=\{0\}$, and the algebra can be reduced 
to a 2-term L$_{\infty}$ 
algebra. Indeed, the action of $v\in V$ on $U$ that is implicit in (\ref{closureEXP}) then reduces to 
 \be
   u \ \mapsto \ v(u) \ \equiv \ -[v,u] \ \in \ U\;. 
 \ee 
Using this and ${\rm Jac}(v_1,v_2,v_3)=f(v_1,v_2,v_3)$, it is straightforward to verify that all products in (\ref{ALLproducts}) that take values in $X_2$ trivialize. 
In particular, $\ell_4$ trivializes. Theorem 1 is contained as a special case, for which $U=V$.

\section{Examples}

We will now discuss a few examples, which get increasingly less trivial, with the goal 
to illustrate the scope of the above theorems. \\[2ex]
\textit{The octonions:} The seven imaginary octonions $e_a$, $a=1,\ldots, 7$ 
satisfy the algebra 
 \be
  e_a e_b \ = \ -\delta_{ab}1 + \eta_{abc} \, e_c\;, 
 \ee
and thus the commutation relations 
 \be\label{l2rela}
  [e_a, e_b] \ = \ 2\,\eta_{abc} \, e_c\;, 
 \ee   
where the structure constants are defined as follows. Splitting the index as 
$a=(i,\bar{i},7)$, where $i,\bar{i}=1,2,3$, $\eta_{abc}$ is the totally antisymmetric tensor 
defined by 
 \be
  \eta_{ijk} \ = \ \epsilon_{ijk}\;, \qquad \eta_{i\bar{j}\bar{k}} \ = \ -\epsilon_{ijk}\;, \qquad 
  \eta_{7i\bar{j}} \ = \ \delta_{ij}\;, 
 \ee 
with the three-dimensional  Levi-Civita symbol satisfying $\epsilon_{123}=1$.  
(This  coincides with the conventions of  \cite{Gunaydin:2016axc}.)   
The $\eta_{abc}$ satisfy the following relations 
 \be\label{importantetaREL}
  \begin{split}
   \eta_{abe}\,\eta_{cde} \ &= \ 2\,\delta_{c[a} \delta_{b]d} - \Theta_{abcd}\;, \\
   \Theta_{abcd} \ &\equiv \ \tfrac{1}{3!}\,\epsilon_{abcdefg}\, \eta_{efg}\;. 
  \end{split}
 \ee  
Using these it is straightforward to compute  
the Jacobiator: 
 \be\label{covJAC}
  {\rm Jac}(e_a,e_b,e_c) \ = \ -12\,\Theta_{abcd}\, e_d\;. 
 \ee
It is easy to verify with this expression that each generator appears on the right-hand side, see 
\cite{Gunaydin:2016axc}. 
Thus, the Jacobiator does not take values in a proper subspace, and therefore the L$_{\infty}$ 
extension requires a doubling to a 14-dimensional space (with basis $\{e_a, e_a^*\}$)  as in Theorem 1, 
with the non-trivial brackets being given in addition to (\ref{l2rela}) 
by 
\be
\begin{split}
  \ell_1(e_a^*) \ &= \ e_a\;, \\
  \ell_2(e_a^*, e^{}_b) \ &= \ 2\, \eta_{abc}\, e_c^*\;, \\
   \ell_3(e_a, e_b, e_c) \ &= \ 12\, \Theta_{abcd}\, e_d^*\;. 
\end{split}
\ee
There is no further non-trivial extension; in particular, this algebra cannot describe a 
non-trivial gauge symmetry in a field theory. \\[2ex]
\textit{The R-flux algebra:}  This algebra, introduced  in \cite{Blumenhagen:2010hj,Lust:2010iy,Blumenhagen:2011ph}, 
 is a contraction of the algebra of imaginary octonions in the following sense \cite{Gunaydin:2016axc}:\footnote{As shown in  \cite{Lust:2017bgx},   the algebra of octonions can be also contracted
 in an analogous way to the magnetic monopole algebra, which is isomorphic to the R-flux algebra upon exchange of position and momentum variables.}
We decompose $e_a=(e_i, f_{{i}}, e_7)$, with $i=1,2,3$, and introduce  a scaling parameter 
$\lambda$ to define 
 \be
  p_i \ \equiv \ -\tfrac{1}{2}\lambda i e_i\;, \qquad 
  x^i \ \equiv \ \tfrac{1}{2}\sqrt{\lambda}i f_i\;, \qquad 
   I \ \equiv \ \tfrac{1}{2}i \lambda^{\frac{3}{2}} e_7\;. 
 \ee
Expressing the algebra  (\ref{l2rela}) now in terms of $x, p, I$ and sending $\lambda\rightarrow 0$ one obtains
the $R$-flux algebra 
 \be\label{Ralgebra}
  [x^i, p_j] \ = \ i \delta^i{}_j I\;, \qquad [x^i,x^j] \ = \ i\epsilon^{ijk} p_k\;, \qquad [p_i, p_j] \ = \ 0\;,
 \ee  
where $I$ is a central element that commutes with everything. 
It is easy to see that the only non-vanishing Jacobiator is 
 \be
  {\rm Jac}(x^i, x^j, x^k) \ = \ 3\,\epsilon^{ijk}\, I\;. 
 \ee 
Thus, the Jacobiator takes values in the one-dimensional subspace spanned by $I$. 
According to the specialization discussed after the proof of Theorem 2, 
we can then define an L$_{\infty}$ structure on $X_1+X_0$, 
where $X_0=\{x^i, p_i, I\}$ and $X_1=\{I^*\}$. In addition to the 2-brackets defined 
by (\ref{Ralgebra}) we have the non-trivial products 
 \be
 \begin{split}
  \ell_1(I^*) \ &= \ I\;, 
 \\
  \ell_3(x^i, x^j, x^k) \ &= \ -3\,\epsilon^{ijk}\, I^*\;. 
 \end{split}
 \ee
\\[3ex]
\textit{The Courant algebroid:} The Courant bracket of generalized geometry or the `C-bracket' 
of double field theory have a non-vanishing Jacobiator. Denoting the arguments of this bracket, i.e., 
the elements of $X_0$, by $\xi_1, \xi_2$, etc., it is given by 
 \be
  {\rm Jac}(\xi_1,\xi_2,\xi_3) \ = \ {\cal D}f(\xi_1, \xi_2, \xi_3)\;, \qquad
  f(\xi_1, \xi_2, \xi_3) \ \equiv \ \tfrac{1}{2} \sum_{\rm anti}\langle [\xi_1, \xi_2],\xi_3\rangle\;, 
 \ee
where $\langle\,, \rangle$ denotes the $O(d,d)$ invariant metric, and 
${\cal D}$ is the exterior 
derivative in generalized geometry or the doubled partial derivative in double field theory. 
The bracket satisfies for a function $\chi$
 \be
  [{\cal D}\chi, \xi] \ = \ -\tfrac{1}{2}{\cal D}\langle {\cal D}\chi, \xi\rangle\;,
 \ee
so that for our current notation we read off with (\ref{closureEXP}) 
 \be
  \xi(\chi) \ = \ -\tfrac{1}{2}\langle {\cal D}\chi, \xi\rangle \;. 
 \ee
It was established by Roytenberg and Weinstein that the Courant algebroid defines a 
2-term L$_{\infty}$ algebra with the highest bracket being $\ell_3$, which is defined by $f$,   
and $X_1$ being the space of functions \cite{roytenberg-weinstein}.   
The space $X_2$ of constants (the kernel of the differential operator ${\cal D}$) 
is not needed as all brackets in (\ref{ALLproducts}) taking values in $X_2$ vanish. 
For instance, $\ell_2$ for two functions $\chi_1, \chi_2\in X_1$ becomes 
 \be
  \ell_2(\chi_1, \chi_2) \ = \ -({\cal D}\chi_1)(\chi_2) -({\cal D}\chi_2)(\chi_1) \ = \ 
  \langle {\cal D}\chi_1, {\cal D}\chi_2\rangle \ = \ 0\;. 
 \ee
In double field theory language this  is zero because of the `strong constraint', 
and it is also one of the axioms of a Courant algebroid (see definition 3.2, axiom 4 in \cite{roytenberg-weinstein}).  
The vanishing of all other products taking values in  $X_2$ can be verified similarly 
using the relations given, for instance, in 
\cite{Hohm:2017pnh}. Thus, the existence of an L$_{\infty}$ structure on the Courant 
algebroid is a  corollary of the more general Theorem 2.

\section{Conclusions} 

We established general theorems about the existence of L$_{\infty}$ algebras for a 
given bracket and discussed possible field theory realizations. 
This includes well-known examples such as the Courant algebroid as 
special cases. Most importantly,  
it then remains to construct explicit examples of algebras  
that obey the conditions of Theorem~2 and that really do use the full structure possible, 
particularly a non-trivial $4$-bracket. This may require identifying a structure that relaxes some 
of the axioms of a Courant algebroid. 

Moreover, it is clear that there will be further 
generalizations of this theorem. For instance, the construction of Theorem~2 could be 
extended by taking the map $\ell_1:X_2\rightarrow X_1$ not to be the inclusion map but rather 
a non-trivial operator that again could have a non-trivial kernel, which in turn would necessitate 
a new space $X_3$ and higher brackets beyond a 4-bracket. 
These may be useful for generalizations of double and 
exceptional field theory \cite{Hohm:2013pua}. 
Indeed, it is to be expected that the gauge structure of exceptional field theory requires L$_{\infty}$
algebras with arbitrarily high brackets \cite{Berman:2012vc}, as is also the case in closed string field theory \cite{Zwiebach:1992ie}. 
Moreover, in order to obtain interesting L$_{\infty}$ algebras with non-trivial field theory realizations, for special 
cases it is instrumental to take an appropriate bracket as starting point. For instance, 
for the E$_{8(8)}$ theory in \cite{Hohm:2014fxa} the naive bracket does not yield a Jacobiator living in the image of 
an appropriate operator (or, equivalently, the naive bracket does not transform covariantly under its own `adjoint' action \cite{Cederwall:2015ica}), 
but rather the vector space has to be suitably enlarged from the beginning, 
leading to a so-called Leibniz-Loday structure \cite{Hohm:2018ybo}.

 \section*{Acknowledgments}

We would like to thank Ralph Blumenhagen, Michael Fuchs, Ezra Getzler, Tom Lada, Martin Rocek, Christian Saemann, Jim Stasheff, Richard Szabo and Barton Zwiebach 
for useful discussions and comments. 
O.H.~is supported by a DFG Heisenberg Fellowship 
of the German Science Foundation (DFG). V.K.~is supported by the Capes- Humboldt Fellowship No.~0079/16-2.  The work of D.L.~is supported by the
ERC Advanced Grant No.~320045 ``Strings and Gravity".

\begin{appendix}

\section{A$_\infty$ and non-associative algebras}

In analogy to the doubling of  vector spaces introduced for the L$_\infty$ realization of Theorem~1 
we will show that every non-associative algebra has a realization as an A$_\infty$ algebra. An A$_\infty$ algebra is a graded vector space $V$ together with a collection $\left\lbrace m_k\, |\, k \in \mathbb{N}\right\rbrace$ of multilinear maps $m_k:\bigotimes^{k} V \rightarrow V$ of internal degree $k-2$ satisfying the following fundamental identity \cite{Lada}
\begin{equation}
\sum_{\lambda=0}^{n-1}\,\sum_{j=1}^{n-\lambda}\, (-1)^{j+\lambda+j\lambda+nj+j(|a_1|+...+|a_{\lambda}|)}\, m_{n-j+1}(a_1,...,a_{\lambda},m_j(a_{\lambda +1},...,a_{\lambda+k}),a_{\lambda+k+1},...,a_{n}) \ = \ 0\;, 
\end{equation}
for every $n\in \mathbb{N}$.
The first four equations read explicitly 
\begin{itemize}
\item $n=1$, ${\rm deg}= -2$:
\begin{equation}
0 \ = \ m_1(m_1(a_1))\;. 
\end{equation}
\item $n=2$, ${\rm deg}= -1$:
\begin{equation}
0 \ = \ -m_1(m_2(a_1,a_2))+m_2(m_1(a_1),a_2)+(-1)^{|a_1|}m_2(a_1,m_1(a_2))\;. 
\end{equation}
\item $n=3$, ${\rm deg}= 0$:
\be
\begin{split}
&0 \ = \ m_1(m_3(a_1,a_2,a_3))\\
&\phantom{=}+m_3(m_1(a_1),a_2,a_3)+(-1)^{|a_1|}m_3(a_1,m_1(a_2),a_3)
 \\ &\phantom{+}+(-1)^{|a_1|+|a_2|}m_3(a_1,a_2,m_1(a_3))
+m_2(m_2(a_1,a_2),a_3)-m_2(a_1,m_2(a_2,a_3))\,.
\end{split}
\ee
\item $n=4$, ${\rm deg}= 1$:
\be
\begin{split}
0 \ = \ &-m_1(m_4(a_1,a_2,a_3,a_4))
+m_4(m_1(a_1),a_2,a_3,a_4)+(-1)^{|a_1|}m_4(a_1,m_1(a_2),a_3,a_4)\\
&+(-1)^{|a_1|+|a_2|}m_4(a_1,a_2,m_1(a_3),a_4) 
+(-1)^{|a_1|+|a_2|+|a_3|}m_4(a_1,a_2,a_3,m_1(a_4))\\
&-m_3(m_2(a_1,a_2),a_3,a_4)+m_3(a_1,m_2(a_2,a_3),a_4)-m_3(a_1,a_2,m_2(a_3,a_4))
 \\
&+m_2(m_3(a_1,a_2,a_3),a_4)+(-1)^{|a_1|}m_2(a_1,m_3(a_2,a_3,a_4))\;. 
\end{split}
\ee
\end{itemize}

Let $(V,\star)$ be a non-associative algebra and $V^{*}$ a vector space isomorphic to $V$ with the isomorphism denoted by $V\ni a\mapsto a^*\in V^*$. The graded vector space of the A$_\infty$ algebra is then defined as
\begin{equation}
X_1 \ = \ V^*\;, \qquad X_0 \ = \ V\;. 
\end{equation}
In addition we define the following products 
\begin{equation}
m_1(a^*) \ = \ a\, , \qquad m_2(a_1, a_2) \ = \ a_1\star a_2\;. 
\end{equation}   
Using this construction, the $n=1$ A$_\infty$ equation is trivially satisfied. For the second equation we compute
\begin{align}
0 \ &= \ -m_1(m_2(a_1^*,a_2))+m_2(m_1(a_1^*),a_2)+(-1)^{|a_1^*|}m_2(a_1^*,m_1(a_2))\\
\ &= \ -m_1(m_2(a_1^*,a_2))+a_1\star a_2 \;, 
\end{align}
from which we conclude 
\begin{align}
m_2(a_1^*,a_2) \ = \ (a_1\star a_2)^*\;. 
\end{align}
For two arguments of degree 1 we compute 
\begin{align}
0 \ &= \ -m_1(m_2(a_1^*,a_2^*))+m_2(m_1(a_1^*),a_2^*)+(-1)^{|a_1^*|}m_2(a_1^*,m_1(a_2^*))\\
\ &= \ m_2(a_1,a_2^*)-(a_1\star a_2)^*\;, 
\end{align}
from which we conclude 
\begin{align} 
m_2(a_1,a_2^*) \ = \ (a_1\star a_2)^* \;. 
\end{align}
Note that the $m$-products have no a priori symmetry properties, so the $m_2$-product 
has to be specified for each order of entries individually. 

The $n=3$ equations read
\begin{align}
0 \ &= \ m_1(m_3(a_1,a_2,a_3))+m_2(m_2(a_1,a_2),a_3)-m_2(a_1,m_2(a_2,a_3))\\
\ &= \ m_1(m_3(a_1,a_2,a_3))+(a_1\star a_2)\star a_3-a_1\star (a_2\star a_3)\;,  
\end{align}
from which we infer that the 3-product is defined by the associator: 
\begin{align}
m_3(a_1,a_2,a_3) \ = \ -{\rm Ass}(a_1,a_2,a_3)^*\;. 
\end{align}  
Moreover, for total degree 1 we compute  
\begin{align}
0 \ &= \ m_3(m_1(a_1^*),a_2,a_3)+m_2(m_2(a_1^*,a_2),a_3)-m_2(a_1^*,m_2(a_2,a_3))\\
\ &= \ -{\rm Ass}(a_1,a_2,a_3)^*+((a_1\star a_2)\star a_3)^*-(a_1\star (a_2\star a_3))^*\;, 
\end{align}
which is therefore satisfied. 

We claim that the  $n=4$ equations are satisfied for $m_4\equiv 0$, which we verify 
by a direct computation: 
\begin{align}
0 \ &= \ -m_3(m_2(a_1,a_2),a_3,a_4)+m_3(a_1,m_2(a_2,a_3),a_4)-m_3(a_1,a_2,m_2(a_3,a_4))\\
&\phantom{=}+m_2(m_3(a_1,a_2,a_3),a_4)+m_2(a_1,m_3(a_2,a_3,a_4) \nonumber \\
\ &= \ {\rm Ass}((a_1\star a_2),a_3,a_4)^*-{\rm Ass}(a_1,a_2\star a_3,a_4)^*+{\rm Ass}(a_1,a_2,a_3\star a_4)^*\nonumber \\
&\phantom{=}-({\rm Ass}(a_1,a_2,a_3) \star a_4)^*-(a_1\star {\rm Ass}(a_2,a_3,a_4))^*\nonumber \\
\ &= \ \bigg[((a_1\star a_2)\star a_3)\star a_4-(a_1\star a_2)\star (a_3\star a_4)-(a_1 \star (a_2 \star a_3))\star a_4+a_1\star ((a_2 \star a_3)\star a_4) \nonumber \\
&\phantom{=}+(a_1\star a_2)\star (a_3\star a_4)-a_1\star (a_2\star (a_3 \star a_4)) - ((a_1\star a_2)\star a_3)\star a_4+(a_1 \star (a_2 \star a_3))\star a_4 \nonumber \\
&\phantom{=}-a_1\star ((a_2 \star a_3)\star a_4)+a_1\star (a_2\star (a_3 \star a_4)) \bigg]^* \nonumber\,. 
\end{align}
The terms exactly cancel. This completes the proof that any non-associative algebra can be embedded into an 
A$_\infty$ algebra.

\end{appendix}

\end{document}